# The fast light of CsI(Na) crystals[*]

Xilei Sun[1;1)]    Junguang Lu[1]    Tao Hu[1]    Li Zhou[1]    Jun Cao[1]    Yifang Wang[1]    Liang Zhan[1]    Boxiang Yu[1]

Xiao Cai[1]    Jian Fang[1]    Yuguang Xie[1]    Zhenghua An[1]    Zhigang Wang[1]    Zhen Xue[1]    Aiwu Zhang[1]

Qiwen Lu[2]    Feipeng Ning[1]    Yongshuai Ge[1]    Yingbiao Liu[1]

[1]Institute of High Energy Physics, CAS, Beijing 100049, China

[2]Shanxi University, Taiyuan 030006, China

**Abstract** The responds of different common alkali halide crystals to alpha-rays and gamma-rays are tested in this research. It is found that only CsI(Na) crystals have significantly different waveforms between alpha and gamma scintillations, while others have not this phenomena. It is suggested that the fast light of CsI(Na) crystals arises from the recombination of free electrons with self-trapped holes of the host crystal CsI. Self-absorption limits the emission of fast light of CsI(Tl) and NaI(Tl) crystals.

## 1 Introduction

CsI(Na) crystals are important scintillators for electromagnetic calorimetry in experimental particle physics. According to catalogues of main scintillation crystals producers (Saint-Gobain, AMCRYS-H et al.), the decay time of γ-scintillations in CsI(Na) amounts to approximately 630 ns with the peak wavelength of 420 nm. In fact, our previous paper reveals that there are fast components hiding in the slow γ-scintillations of CsI(Na) crystals, and the proportion of that is larger responding to higher dE/dx particles [1].

In this paper, we report the result of the testing of other common alkali halide crystals responding to alpha-rays and gamma-rays. The luminescence process of common alkali halide crystals is discussed in detail at the end of this paper.

## 2 Experimental set-up

The schematic of the experimental set-up is illustrated in Fig. 1. Two R8778 PMTs are directly attached to the top and the bottom surfaces of the crystal. The PMT signals are sent to the oscilloscope and discriminators (NIM CAEN N840) via a fan-out module (NIM CAEN N625). The discriminator thresholds are set to be equivalent to 0.5 single photoelectrons, which are calibrated by a pulsed LED light source. The noises of PMT can then be effectively suppressed to 0.055 Hz, which is mainly caused by Cherenkov light, by the coincidence of the two discriminator signals with a width of 200 ns. The oscilloscope is a TDS3054C with a sampling frequency of 5 GS/s and a memory depth of 2 μs (10000 points, 0.2 ns/channel) for each of 4 channels. The high-speed sampling rate is a necessary condition for fast waveform analysis.

[*] Supported by Technological Innovation Project of Institute of High Energy Physics

[1)] E-mail: sunxl@ihep.ac.cn

The crystals tested in this study are CsI(Na), CsI(Tl), NaI(Tl) and pure CsI. The detail properties of them are listed in Table 1.

Gamma-rays from a 0.5 micro-Curie $^{137}$Cs source and alpha-rays from a 5 micro-Curie $^{239}$Pu source are used to excite the crystals.

Table. 1 The crystals tested in the research.

| Crystals | CsI(Na) | CsI(Tl) | NaI(Tl) | CsI |
|---|---|---|---|---|
| Doping | Na | Tl | Tl | - |
| Mole% | ~ 0.02 | ~ 0.2 | ~ 0.2 | - |
| Size (cm) | 2.5×2.5×2.5 | Φ 2.5×2.5 | 2×2×2 | Φ 2.5×2.5 |

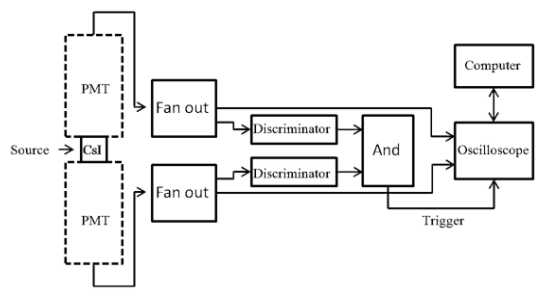

Fig. 1. Experimental set-up for the test of the crystals.

## 3 Results and discussion

The measured waveform profile histograms of γ-scintillations and α-scintillations from different crystals are shown respectively below. Here we chose waveforms of full energy deposition to build the histograms, which are 661.7 keV for gamma and 5.2 MeV for alpha. The values of the rise time are read from the amplification plots of the leading edge directly, and that of decay time are get from an exponential fitting of the trailing edge.

### 3.1 Waveforms of CsI(Na)

Waveform profile histograms of γ-scintillations (661.7 keV) and α-scintillations (5.2 MeV) from CsI(Na) crystals are shown in Fig. 2. The waveforms of γ-scintillations have a rise time of ~ 40 ns and a decay time of ~ 670 ns. In comparison, the waveforms of α-scintillations with old or new surface have significant fast components. The decay time is only ~ 17 ns and the rise time is ~ 5 ns. The decay time of slow components of α-scintillations is shorter than that of γ-scintillations, which is ~ 490 ns.

Here, the old surface is the surface layer of CsI(Na) crystals with sodium ions losing caused by the moisture absorption of sodium; the new surface is that of grinding off the old layer. Thallium-doped crystals do not have this phenomenon.

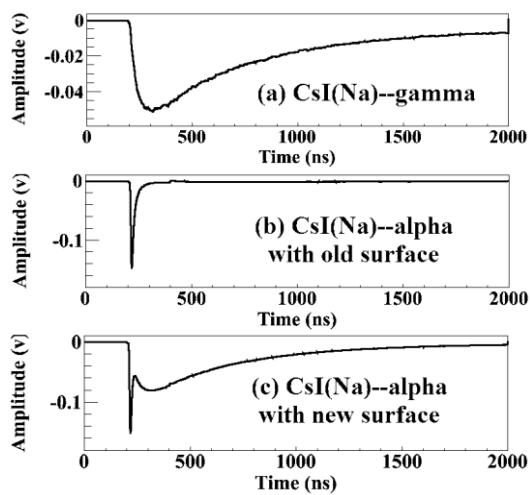

Fig. 2 Waveform profile histograms from CsI(Na) crystals. (a) γ-scintillations of 661.7 keV (b) α-scintillations of 5.2 MeV with old surface (c) α-scintillations of 5.2 MeV with new surface.

### 3.2 Waveforms of CsI(Tl)

Waveform profile histograms of γ-scintillations (661.7 keV) and α-scintillations (5.2 MeV) from CsI(Tl) crystals are shown in Fig. 3. The rise time of α-scintillations is ~ 35 ns and the decay time is ~ 670 ns. There is not fast light. Similar, the waveforms of γ-scintillations have a rise time ~ 50 ns and a longer decay time ~ 1080 ns. Although the different decay time can be used for alpha-gamma separation, but the performance is much less than CsI(Na) crystals by using the fast light.

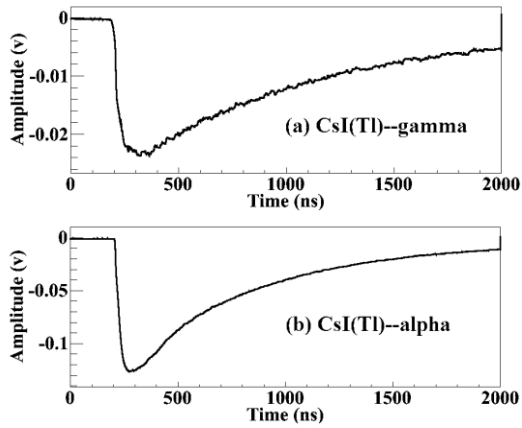

Fig. 3 Waveform profile histograms from CsI(Tl) crystals. (a) γ-scintillations of 661.7 keV (b) α-scintillations of 5.2 MeV.

### 3.3 Waveforms of NaI(Tl)

Waveform profile histograms of γ-scintillations (661.7 keV) and α-scintillations (5.2 MeV) from NaI(Tl) crystals are shown in Fig. 4. The rise time of α-scintillations is ~ 10 ns and the decay time is ~ 170 ns. Similar, the waveforms of γ-scintillations have a rise time ~ 20 ns and a slightly longer decay time ~ 250 ns. It is difficult to separate alpha and gamma by the decay time.

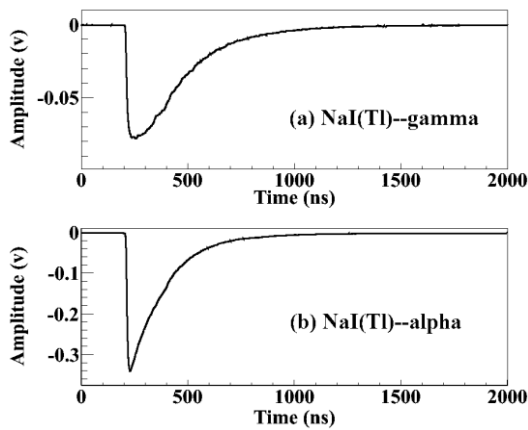

Fig. 4 Waveform profile histograms from NaI(Tl) crystals. (a) γ-scintillations of 661.7 keV (b) α-scintillations of 5.2 MeV.

### 3.4 Waveforms of pure CsI

Finally, waveform profile histograms of γ-scintillations (661.7 keV) and α-scintillations (5.2 MeV) from pure CsI crystals are shown in Fig. 5. The rise time of α-scintillations is ~ 5 ns and the decay time is ~ 15 ns. Similar, the waveforms of γ-scintillations have a rise time ~ 5 ns and a decay time ~ 22 ns.

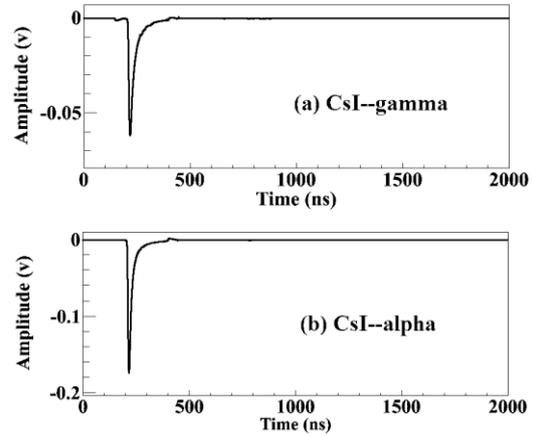

Fig. 5 Waveform profile histograms from pure CsI crystals. (a) γ-scintillations of 661.7 keV (b) α-scintillations of 5.2 MeV.

The summary of the measured waveform parameters are listed in Table 2. The present measurements of waveforms lead to two facts: (1) The fast light of CsI(Na) crystals can be significantly excited by alpha-rays, while which cannot be excited by gamma-rays. (2) The luminescence of gamma and alpha from the other common alkali halide crystals is not significantly different. It, thus, appears from these two facts that there are different aspects of the luminescence mechanisms between CsI(Na) crystals and CsI(Tl) or NaI(Tl) crystals.

Table. 2 Summary of waveform parameters.

| Crystals | CsI(Na) | | CsI(Tl) | | NaI(Tl) | | CsI | |
|---|---|---|---|---|---|---|---|---|
| Source | γ | α | γ | α | γ | α | γ | α |
| Rise time (ns) | 40 | 5 | 50 | 35 | 20 | 10 | 5 | 5 |
| Decay time (ns) | 670 | 17$_{fast}$ | 1080 | 670 | 250 | 170 | 22 | 15 |
| | | 490$_{slow}$ | | | | | | |

### 3.5 Discussion

The scintillation process in common alkali halide crystals has been discussed by Stephen E. Derenzo and Marvin J. Weber in reference [2], which can be described in several steps:

**i. Ionization**

Ionization occurs when the particles injected into the crystals, and the average energy needed to produce an electron-hole pair is about 20 eV for CsI and NaI crystals.

**ii. Relaxation of Electrons and Holes**

The electrons and the holes resulting from the ionization process form separated charge carriers that cause the surrounding atoms to rearrange themselves. In CsI and NaI crystals, the self-trapped hole takes the form of a so-called $V_k$ center, where two Iodine atoms share the hole by pulling together and forming a covalent bond.

**iii. Carrier Diffusion**

In CsI and NaI crystals, the valance bands are filled and an excess electron is spatially diffuse in an essentially empty conduction band. On the other hand, holes are generally localized and will only diffuse if thermal vibrations are able to move the hole from one trapping site to another.

**iv. Formation of different excitons**

A diffuse electron may be trapped by an activator atom and then a self-trapped hole also be captured by it through thermal migration. The result is an activator-trapped-exciton. A diffuse electron may also be trapped immediately by a self-trapped hole resulting a self-trapped-exciton. The probability that a given electron will recombine with a self-trapped hole rather than suffer capture at activator atom site will be an increasing function of the density of self-trapped holes, hence, an increasing function of ionization density, while a decreasing function of concentration of activator atoms.

**v. Radiative Emission**

The de-excitation of two different kinds of excitons results two different scintillations. Thermal quenching is a common non-radiative process and occurs if thermal vibrations can deform exciton to ground state. Self-absorption is another constraint of scintillations if the emission band overlaps with the absorption band.

Based on the above process, it is noted that the fast scintillations with decay time of about 20 ns and wavelength of about 310 nm from pure CsI or NaI crystals are dominated by self-trapped-excitons for both gamma and alpha particles. The fast light of α-scintillations from CsI(Na) crystals arises from self-trapped-excitons, while slow components of both alpha and gamma scintillations are dominated by activator-trapped-excitons. Fortunately, the absorption bands of CsI(Na) are all below 300 nm [3], thus, the fast light can be propagation through crystals.

CsI(Tl) and NaI(Tl) crystals have the same mechanism of scintillation, while, a discrepancy is that the emission bands of fast light overlaps with the absorption bands [4] [5], hence, the fast light of high ionization density particles cannot be significantly observed. UV self-absorption of CsI(Tl) and NaI(Tl) crystals may be due to the presence of Tl atom, which can be excited by the UV light.

The transmittance of different crystals measured by a DH-2000-BAL Deuterium-Halogen Light Source and a QE65000 Scientific-grade Spectrometer is shown in Fig. 6. It can clearly be seen that the light of wavelength less than 320nm is strongly absorbed for CsI(Tl) crystals. The decay length of 310nm light is ~ 33 cm for CsI(Na) crystals according to the measurements.

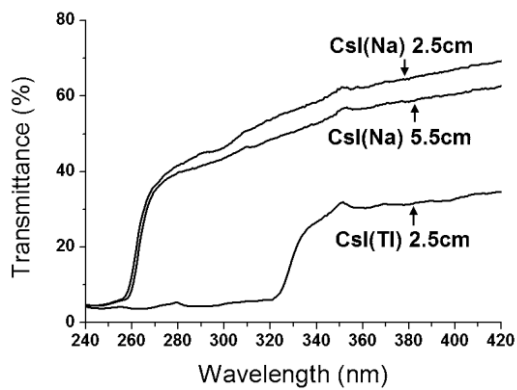

Fig. 6 Transmittance of different crystals.

## 4 Conclusions

Different alkali halide crystals are tested, the fast components of α-scintillations only can be significantly found from CsI(Na) crystals. Self-absorption limits the emission of fast light from CsI(Tl) and NaI(Tl) crystals. Hence, CsI(Na) is the unique crystals among common alkali halide crystals, which can be used to doing particle identification by different kinds of luminescence mechanisms.

The fast light of CsI(Na) crystals can be significantly excited by particles with high ionization density like alphas rather than electrons. As a result, the high dE/dx particles can be easily distinguished from gamma-rays and electrons. Therefore, CsI(Na) crystals are suitable materials for neutron detection and dark matter detection.

## 5 Acknowledgments

We are thankful to Liangjian Wen for his support.